\documentclass[twocolumn,showpacs,preprintnumbers,amsmath,amssymb]{revtex4}
\newcommand\beq{\begin{equation}}
\newcommand\eeq{\end{equation}}

\usepackage{graphicx}% Include figure files
\usepackage{dcolumn}% Align table columns on decimal point
\usepackage{bm}% bold math
\usepackage{amsmath}
\usepackage{amsfonts}   % if you want the fonts
\usepackage{amssymb}

\begin{document}

\preprint{LIGO-P070066-01-Z}

%%%%%%%%%%%%%%%%%%%%%      created 14/10/2006  %%%%%%%%%%%%%%%%%%%%%
%%%%%%%%%%%%%%%%%%%%%      last revised 03/07/2007  %%%%%%%%%%%%%%%%%%%%%

\title{Perspectives on Beam-Shaping Optimization for Thermal-Noise Reduction in Advanced Gravitational-Wave Interferometric Detectors: Bounds, Profiles, and Critical Parameters}% Force line breaks with \\

\author{Vincenzo Pierro}%
\author{Vincenzo Galdi}\email{vgaldi@unisannio.it} \homepage{http://www.ing.unisannio.it/vgaldi}
\author{Giuseppe Castaldi}%
\author{Innocenzo M. Pinto}%
\affiliation{Waves Group, Department of Engineering, University of Sannio, I-82100 Benevento, Italy\\
}%

\author{Juri Agresti}
\author{Riccardo DeSalvo}
\affiliation{LIGO Laboratory, California Institute of Technology, Pasadena, CA 91125, USA}%
\date{\today}% It is always \today, today,

\begin{abstract}
Suitable shaping (in particular, {\em flattening} and {\em broadening}) of the laser beam has recently been proposed as an effective device to reduce internal (mirror) thermal noise in advanced gravitational wave interferometric detectors.
Based on some recently published analytic approximations (valid in the infinite-test-mass limit) for the Brownian and thermoelastic mirror noises in the presence of {\em arbitrary-shaped} beams, this paper addresses certain preliminary issues related to the {\em optimal beam-shaping} problem. In particular, with specific reference to the Laser Interferometer Gravitational-wave Observatory (LIGO) experiment, absolute and realistic {\em lower-bounds} for the various thermal noise constituents are obtained and compared with the current status (Gaussian beams) and trends (``mesa'' beams), indicating fairly ample margins for further reduction. In this framework, the effective dimension of the related optimization problem, and its relationship to the critical design parameters are identified, physical-feasibility and model-consistency issues are considered, and possible
additional requirements and/or prior information exploitable to drive the subsequent optimization process are highlighted.
\end{abstract}

\maketitle

%%%%%%%%%%%%%%%%%%%%%%%%%%%%%%%%%%%%%%%%%%%%%%%%%%%%%%%%%%%%%%%%%
\section{Introduction}
%%%%%%%%%%%%%%%%%%%%%%%%%%%%%%%%%%%%%%%%%%%%%%%%%%%%%%%%%%%%%%%%%
\label{Intro}
In all currently operating (and possibly future) interferometric gravitational wave detectors, the overall limit sensitivity of the instrument is bounded by the noise floor, which, in the most interesting observational frequency band (30--300 Hz), is dominated by thermal noises in the substrate and in the high-reflectivity coating of the test masses. With particular reference to the Laser Interferometer Gravitational-wave Observatory (LIGO) experiment \cite{LIGOweb}, an introductory discussion of the various noise components can be found in \cite{Rao_thesis}, and a numerical code for computing the noise budget is available from \cite{Bench}. Toward the development of second-generation detectors, such as Adv-LIGO \cite{AdvLIGO}, the quest for increasing the event rate in the observational band has motivated the exploration of various techniques for reducing the mirror thermal noise. With specific reference to the coating Brownian noise (dominant in the current baseline design featuring fused-silica test-masses), use of improved (low-mechanical-loss) materials \cite{LMA}, geometric optimization of the coating design \cite{SPIE2006}, and {\em flat-top} (commonly referred to as ``mesa'') beams \cite{Thorne2000b,Dambrosio2004a} seem the most promising. The latter option, intuitively motivated by the potential capability of a mesa beam (MB) of better averaging the thermally-induced mirror surface fluctuations as compared to a standard Gaussian beam (GB), has been numerically proved to yield significant reductions in the overall thermal noise \cite{Oshaugh2004, Agresti2005b}, and has led to the development of a cavity prototype with non-spherical ``Mexican hat'' (MH) profile mirrors \cite{Agresti2005,Beyer}. Alternative (nearly-concentric \cite{Savov2006}, nearly-spheroidal \cite{Bondarescu2006,Galdi2006,Galdi2006a}) designs have been subsequently proposed to cope with the inherent tilt-instability of the originally-conceived nearly-flat configuration. Also, use of higher-order modes in standard spherical cavities has been shown to provide, in principle, comparable reductions without the need of changing the mirror profile \cite{Vinet2}, but its practical feasibility still remains to be assessed.

The method utilized in \cite{Agresti2005b} to compute the coating and substrate thermal noises relies on a finite-test-mass (FTM) computationally-intensive numerical analysis based on the approach in \cite{Bondu,LiuThorne}. More recently \cite{OShaug2006,Lovelace2006}, 
a general though simple formula has been derived in the infinite-test-mass (ITM) limit, which allows the computation of the above noises for {\em arbitrary-shaped} beams. This approximation has been validated and calibrated in \cite{Lovelace2006} against the FTM numerical solutions (see also the discussion in Sec. \ref{Sect:ITMValidity}). In view of its remarkably simple form, in terms of spectral integral functionals of the beam intensity profile, it appears suggestive to exploit it for addressing the {\em optimal beam-shaping} problem, i.e., finding the beam profile that minimizes a given thermal noise constituent. In a step-by-step approach, acknowledging the formal and computational complexity of the arising optimization problem, this paper addresses some key preliminary issues. In particular, emphasis is placed on the {\em a priori} deduction of absolute and realistic lower-bounds for the various thermal noise constituents, the identification of
the effective dimension of the problem, and how this depends on the critical design parameters, and the gathering of 
additional requirements and/or prior information to be utilized in the actual optimization problem.  

Accordingly, this paper is organized as follows. In Sect. \ref{Problem}, the problem geometry, formulation, notation and strategy are outlined, with a compact review of the relevant background theory (ITM approximation). In Sect. \ref{Sect:theoret}, under the idealized assumption of
zero diffraction-loss ({\em compact spatial support}) beam profiles, absolute lower bounds for the 
noise constituents, as well as the corresponding beam profiles over the mirror, are obtained in analytic form, by solving a straightforward variational problem. Subsequently, a key physical-feasibility constraint (related to the finite spatial bandwidth of the cavity eigenmodes) is taken into account by approximating the above compact-support optimal profiles in a suitable $L^2$ functional subspace, whose dimension is fixed by the diffraction-loss constraint. This results in more realistic {\em tighter} bounds. In this framework, the role of the number of {\em electromagnetic degrees of freedom} \cite{France} of the cavity in setting the effective dimension of the optimization problem is highlighted. In Sect. \ref{Sect:current}, the obtained absolute and realistic lower bounds for the considered noise components are compared to the levels currently achievable using GB and MB profiles. Moreover, some model-consistency issues are discussed in order to assess the practical relevance of the results.
Finally, in Sect. \ref{Conclusions}, conclusions and recommendations are provided.

%%%%%%%%%%%%%%%%%%%%%%%%%%%%%%%%%%%%%%%%%%%%%%%%%%%%%%%%%%%%%%%%%
\section{Problem Statement}
%%%%%%%%%%%%%%%%%%%%%%%%%%%%%%%%%%%%%%%%%%%%%%%%%%%%%%%%%%%%%%%%%
\label{Problem}

%================================================================
\subsection{Geometry}
\begin{figure}
\begin{center}
\includegraphics[width=8cm]{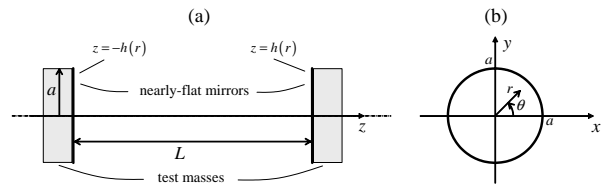}
\end{center}
\caption{Problem schematic: A perfectly symmetric Fabry-Perot optical cavity composed of two nearly-flat mirrors (with profile $h(r)$) attached on cylindrical test masses of radius $a$ separated by a distance $L$ along the z-axis. (a): Side view. (b): Front view.}
\label{Figure1}
\end{figure}

%================================================================
Referring to the problem geometry illustrated in Fig. \ref{Figure1}, we consider a standard Fabry-Perot optical cavity with two identical, symmetric, {\em nearly-flat} (nonspherical) mirrors of radius $a$ laid on cylindrical test masses, separated by a distance $L$ (see Fig. \ref{Figure1}a). The mirror (axisymmetric) departure from flatness is described by $h(r)$, with $r$ denoting the radial coordinate in the mirror plane (see Fig. \ref{Figure1}b). In what follows, attention is focused on the axisymmetric (i.e., $\theta$-independent) eigenmode field distribution $\Phi(r)$ on the mirror, with implicit $\exp(\imath\omega t)$ time-harmonic dependence.
Note that, in view of the duality relations expounded in \cite{Agresti2005a}, the results derived hereafter apply to the {\em nearly-concentric} case too \footnote{This investigation does not deal with tilt (in)stability issues, for which nearly-flat and nearly-concentric configurations exhibit significantly different responses \cite{Savov2006}.}.

%############################################################
%                Figure1
%
%###########################################################

%================================================================
\subsection{Background: Infinite-Test-Mass Approximations}
%================================================================

In the ITM approximation \cite{OShaug2006,Lovelace2006}, and in the low frequency limit of interest
for gravitational-wave interferometers, the power spectral densities of the main coating  and substrate thermal noise constituents of interest can be written as
\beq
S=C \int_{0}^{\infty} {\kappa}^{q+1} \left\{{\cal H}\left[\left|\Phi\right|^2\right]({\kappa})\right\}^2 d{\kappa},
\label{eq:PSD}
\eeq
where $C$ is a noise-type- and frequency-dependent factor (irrelevant for all further developments), $q$ is a noise-type-dependent scaling exponent (see Table \ref{Table1}), $\Phi(r)$ is the axisymmetric eigenmode field distribution on the mirror, and
\beq
{\cal H}[F]({\xi})\equiv\int_{0}^{\infty} F(\zeta) J_{0}(\xi \zeta) \zeta d\zeta
\eeq
denotes the Hankel-transform (HT) operator. 
Here and henceforth, $J_m$ denotes an $m$th-order Bessel function of the first kind \cite[Sec. 9.1]{Abramowitz}.
The (axisymmetric) field distribution $\Phi(r)$ satisfies the eigenvalue equation \cite{Siegman}
\beq
\gamma\Phi(r)=\int_{0}^{a}K(r,r')\Phi(r')r'dr',
\label{eq:integral}
\eeq
where $\gamma$ denotes the half-roundtrip eigenvalue, and the kernel is given by
\begin{eqnarray}
K(r,r')&=&\frac{\imath k}{L}J_0\left(\frac{krr'}{L}\right)\exp\left(-\imath k L\right)\nonumber\\
&\times&\exp\left\{\imath k\left[h(r)+h(r')-\frac{(r^2+r^{\prime 2})}{2L}\right]\right\},
\label{eq:kernel}
\end{eqnarray}
with $k=2\pi/\lambda$ denoting the free-space wavenumber ($\lambda$ being the wavelength). 
Equations (\ref{eq:integral}) and (\ref{eq:kernel}) can be recognized as a mapping between a mirror profile $h(r)$ and a set  $\Omega(h)=\{ [
\gamma_m,\Phi_m], m=1,2,\dots\}$ of eigenstates.
Here and henceforth,
unless otherwise specified, the field distribution on the mirror is assumed to be normalized as follows
\beq
\int_{0}^{\infty} \left|\Phi(r)\right|^2 r dr=1.
\label{eq:norm}
\eeq
In addition, a further constraint has to be enforced on the diffraction loss \cite{Siegman}
\beq
{\cal L}[\Phi]\equiv\int_{a}^{\infty} \left|\Phi(r)\right|^2 r dr=1-\left|{\gamma}\right|^2\le {\cal L}_T,
\label{eq:diffloss}
\eeq
with ${\cal L}_T$ denoting a design limiting value. For Adv-LIGO, the reference figure is ${\cal L}_T=1$ppm ($10^{-6}$). 
The diffraction-loss constraint singles out a subset $\Omega_C(h) \subset \Omega(h)$  of  {\em admissible} eigenmodes.

%%%%%%%%%%%%%%%%%%%%%%%%%%%%%%%%%%%%%%%%%%%%%%%%%%%%%%%
\begin{table}
\caption{Thermal noise constituents of interest and corresponding scaling exponents (cf. (\ref{eq:PSD})).}
\begin{ruledtabular}
\begin{tabular}{lc}                                                  
		Noise type&  $q$  \\
		\hline
		Substrate Brownian & -1\\
		Substrate thermoelastic & 1\\
		Coating Brownian and thermoelastic & 0\\
\end{tabular}
\end{ruledtabular}
\label{Table1}
\end{table}
%%%%%%%%%%%%%%%%%%%%%%%%%%%%%%%%%%%%%%%%%%%%%%%%%%%%%%%

%================================================================
\subsection{Formulation and Notation}
%================================================================
\label{FormNot}
It is expedient to recast the problem into a canonical form by introducing the scaled variables
\beq
{\bar r}=\frac{r}{a},~~{\bar {\kappa}}=a {\kappa},
\eeq
and the scaled field distribution
\beq
\phi({\bar r})=a\Phi({\bar r}a).
\label{eq:phi}
\eeq
Here and henceforth, the overbar denotes scaled quantities.
The noise functional in (\ref{eq:PSD}) can accordingly be rewritten as
\beq
S=\frac{C}{a^{q+2}} {\bar S}\left[\left|\phi\right|^2,q\right],
\label{eq:caN_Dorm}
\eeq
where
\beq
{\bar S}\left[\left|\phi\right|^2,q\right]=\int_{0}^{\infty} {\bar \kappa}^{q+1} \left\{{\cal H}\left[\left|\phi\right|^2\right]({\bar \kappa})\right\}^2 d{\bar \kappa}, 
\label{eq:caN_Dorm1}
\eeq
thereby explicitly factoring out the $a^{-(q+2)}$ scaling law predicted by the ITM approximation \cite{OShaug2006,Lovelace2006}. In what follows, we focus on the scaled noise functional in (\ref{eq:caN_Dorm1}), which essentially accounts for the beam-shaping effects. Unless strictly needed, the explicit dependence on $\left|\phi\right|^2$ and $q$ will be omitted for simplicity of notation.
The scaled field distribution $\phi({\bar r})$ in (\ref{eq:phi}) satisfies the scaled version of the eigenproblem in (\ref{eq:integral}), which can be conveniently recast as
\beq
{\bar \gamma} \phi({\bar r})=\imath \pi N_D \exp\left[-\imath V({\bar r})\right] {\cal H}_1\left[\exp\left(-\imath V\right)\phi\right](\pi N_D {\bar r})
\label{eq:scaled_eigp},
\eeq
where ${\bar \gamma}=\gamma \exp\left(\imath kL\right)$,
\beq
{\cal H}_1[F]({\xi})\equiv\int_{0}^{1} F(\zeta) J_{0}(\xi \zeta) \zeta d\zeta
\label{eq:H1}
\eeq
denotes the $[0,1]$ interval-windowed HT operator, and
\beq
V({\bar r})=k h(a{\bar r})-\frac{\pi N_D {\bar r}^2}{2}
\label{eq:pf}
\eeq
is a mirror-profile-dependent phase function, with
\beq
N_D\equiv2 N_F=\frac{2a^2}{\lambda L}
\label{eq:EDF}
\eeq
denoting twice the Fresnel number $N_F$ of the optical cavity \cite{Siegman}. Following \cite{France}, we shall refer to $N_D$ in (\ref{eq:EDF}) as
 the number of {\em electromagnetic degrees of freedom} \footnote{The concept of electromagnetic degrees of freedom \cite{France} is related to the minimum number of basis functions that are needed to represent within a bounded region, with a prescribed error in a given metric, an electromagnetic field produced by finite-support (impressed or equivalent) sources.}, whose relevance will be illustrated later on (see Sec. \ref{Sect:TighterBounds}).
In the following we shall always assume the eigenfunctions as normalized, viz., 
\beq
\left\|
\phi
\right\|\equiv
\left[\int_{0}^{\infty} \left|\phi({\bar r})\right|^2 {\bar r} d{\bar r}\right]^{\frac{1}{2}}
=1,
\label{eq:norm1}
\eeq
with $\left\|  \cdot \right\|$  denoting the usual 
$L^2_{[0,\infty[}$ (cylindrical) Hilbert norm. Accordingly, we shall write the diffraction-loss constraint as
\beq
{\cal L}[\phi]=\int_{1}^{\infty}\left|\phi({\bar r})\right|^2 {\bar r} d{\bar r}=1-\left|{\bar \gamma}\right|^2\le {\cal L}_T.
\label{eq:diffloss1}
\eeq

%================================================================
\subsection{The Optimization Problem}
%================================================================
The optimization problem of interest consists of minimizing the scaled noise functional in (\ref{eq:caN_Dorm1}), acting on the mirror profile $h({\bar r})$, i.e., in finding the special mirror profile $h({\bar r})$ (within a suitable functional class, e.g., $C^\infty$)  for which 
\beq
\min_{\phi\in\Omega_C(h)} 
\left\|
{\bar \kappa}^{\frac{q}{2}}~
{\cal H}\left[\left|\phi\right|^2\right]
\right\|^2
\label{eq:MinProb}
\eeq
takes on  its smallest value,
 $\Omega_C(h)$ denoting 
the subset of eigenmodes obeying
the diffraction-loss constraint
(\ref{eq:diffloss1}). 
The minimization of (\ref{eq:MinProb}), subject to (\ref{eq:diffloss1}), represents a formidable optimization problem, whose {\em well-posedness} (i.e., existence and uniqueness of the solution, and its continuous dependence on data) cannot be taken for granted, 
with the consequent {\em ill-conditioning} problems that may arise in the numerical implementation. A further complication is posed by the general {\em non-convexity} of the problem, which may result in multiple local minima that may trap standard descent-based optimization techniques (e.g., conjugate gradient \cite{Pedregal}) into false solutions. Therefore, {\em global} optimization techniques need to be applied, such as genetic \cite{Rahmat}, evolutionary \cite{Hoorfar}, or particle-swarm \cite{PSO} algorithms, whose convergence is typically rather slow. Taking into account that each iteration in the optimization procedure may require several numerical solutions of the eigenproblem in (\ref{eq:scaled_eigp}), the resulting overall computational burden can become prohibitive. From the above considerations, it is clear that any {\em blind} attempt of attacking such a complex and computationally-demanding problem may be deemed to failure. In a step-by-step approach, it appears more reasonable to start addressing some preliminary issues, such as: 

\begin{itemize}
	\item[{\em i)}] {\em A priori} estimation of realistic {\em lower bounds} for the various noise constituents, and comparison with the current status and trends, in order to assess the potential reduction achievable by further optimization (and, hence, its worthiness). 
	\item[{\em ii)}] Identification of the effective problem's dimension,
as a function of the key cavity design parameters. 
	\item[{\em iii)}] Gathering of {\em prior information} (e.g., optimal beam profiles and associated structural features) to be exploited in order to intelligently drive the optimization process. 
\end{itemize}

The rest of the present paper accordingly deals with the above issues.

%%%%%%%%%%%%%%%%%%%%%%%%%%%%%%%%%%%%%%%%%%%%%%%%%%%%%%%%%%%%%%%%%
\section{Some Theoretical Bounds}
%%%%%%%%%%%%%%%%%%%%%%%%%%%%%%%%%%%%%%%%%%%%%%%%%%%%%%%%%%%%%%%%%
\label{Sect:theoret}

%================================================================
\subsection{Absolute Bounds: Compact-Support Beams}
%================================================================

%----------------------------------------------------------------
\subsubsection{Variational Problem}
\begin{figure}
\begin{center}
\includegraphics[width=8cm]{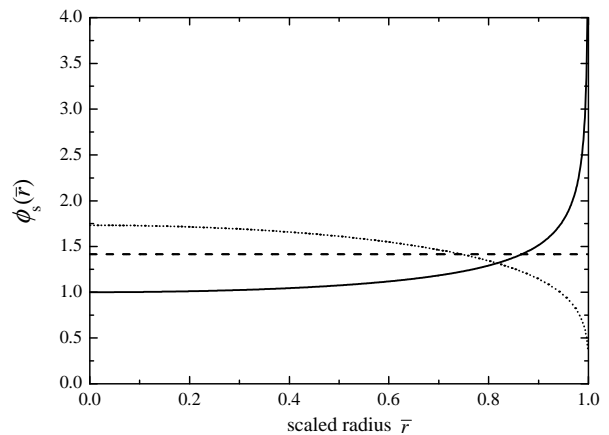}
\end{center}
\caption{Optimal (minimum-noise) compact-support beam profiles in (\ref{eq:fs}). Continuous curve: substrate Brownian ($q=-1$); dashed curve: coating ($q=0$); dotted curve: substrate thermoelastic ($q=1$).}
\label{Figure2}
\end{figure}

%----------------------------------------------------------------
\label{Finite}
The simplest and crudest approximation of the original optimization problem in (\ref{eq:MinProb}), leading to a {\em well-posed} and {\em analytically-treatable} problem, consists of assuming the beam profile to exhibit a {\em compact} spatial support within the $[0,1]$ interval, thereby implying {\em zero diffraction-losses}. As we shall see, this implicitly prevents the profile from being a solution of the eigenproblem in (\ref{eq:scaled_eigp}). Letting $f\equiv\left|\phi\right|^2$, one is thus led to the variational problem in the space $L^1_{[0,1]}$ of summable functions,
\beq
{\bar S}_{abs}^{(min)}\equiv\min_{f\in L^1_{[0,1]}} 
\left\|
{\bar \kappa}^{\frac{q}{2}}~
{\cal H}_1\left[f\right]
\right\|^2,
\label{eq:probcan}
\eeq
under the constraints
\begin{subequations}
\beq
f:[0,1]\rightarrow{\mathbb R}^+,
\label{eq:const1}
\eeq
\beq
\int_0^1 f\left({\bar r}\right) {\bar r} d{\bar r}=1,
\label{eq:const2}
\eeq
\label{eq:const12}
\end{subequations}
whose solution is given below. The arising results are anticipated to provide {\em absolute} lower bounds,
which may {\em not} be attainable, in view of the mentioned unphysical simplifying assumptions.
From the Lagrange theory of constrained optimization \cite{Variational}, the constrained variational problem in (\ref{eq:probcan}) and (\ref{eq:const2}) can be recast into the {\em unconstrained} optimization of the Lagrangian functional \footnote{The satisfaction of the positivity constraint in (\ref{eq:const1}), not explicitly enforced, is verified {\em a posteriori}.}
\beq
\Lambda[f,\mu]=\left\|
{\bar \kappa}^{\frac{q}{2}}~
{\cal H}_1\left[f\right]
\right\|^2
 -2\mu\left[\int_0^1 f\left({\bar r}\right) {\bar r} d{\bar r} -1\right],
\label{eq:lagrangian}
\eeq
where $\mu$ is the so-called Lagrange multiplier. 
It is shown in Appendix \ref{AppA} that this problem admits a {\em unique} solution $f_s$ (i.e., {\em an absolute minimum}), obtainable using variational calculus, viz.,
\beq
f_s\left({\bar r}\right)=\left|\phi_s\left({\bar r}\right)\right|^2=(q+2) (1-{\bar r}^2)^{\frac{q}{2}},~-1\le q\le 1,
\label{eq:fs}
\eeq
which also satisfies the positivity constraint in (\ref{eq:const1}). The corresponding (minimum) noise components are given by
\beq
{\bar S}^{(min)}_{abs}=2^{q+1} \Gamma \left(\frac{q}{2}+1\right) \Gamma \left(\frac{q}{2}+2\right),
\label{eq:theoret}
\eeq
with $\Gamma$ denoting the Gamma (factorial) function \cite[Sec. 6.1]{Abramowitz}. 
The optimal beam profiles are shown in Fig. \ref{Figure2}, whereas the corresponding (minimum) noise values are collected in Table \ref{Table2}. 
%############################################################
%                Figure2
%
%###########################################################

The following remarks are in order:

\begin{itemize}
\item{The noise-minimizing beam profiles can exhibit step discontinuities, or even singularities at ${\bar r}=1$ (see Fig. \ref{Figure2}). This is neither surprising (in view of the relaxation of the physical feasibility constraints) nor undermining of the meaningfulness of the preliminary results derived at this stage as (anticipated) absolute lower bounds for the actual problem. The reader is referred to Sect. \ref{Sect:TighterBounds} below for more {\em realistic} bounds.}

\item{For the coating noises ($q=0$), the optimal profile is {\em perfectly flat}, thereby supporting previous intuitive arguments in favor of flat-top beams \cite{Thorne2000b,Dambrosio2004a}.}

\item{For the substrate noises, the optimal beam profile is appreciably rounded (non-flat) for the 
thermoelastic component ($q=1$). This should be taken into account when assessing the performance of configurations featuring sapphire test-masses, for which substrate thermoelastic noise is known to be dominant \cite{Agresti2005b}. In this framework, use of {\em hyperboloidal} beams \cite{Bondarescu2006,Galdi2006,Galdi2006a} as physically-feasible approximants should be explored. On the other hand, the optimal intensity profile for the Brownian component ($q=-1$) is close to flat, with a steep increase at the mirror's edge. This is clearly unphysical, but may be suggestive of using an annular beam.
Note that the above results pertain to the minimization of a {\em single} noise constituent. 
Extensions to the minimization of a given {\em combination} of noise constituents are possible, but most likely need to be pursued {\em numerically}, via suitable discretization of the involved operators.
}

\end{itemize}

%================================================================
\subsection{More Realistic (Tighter) Bounds: Diffraction-Loss vs. Band-Limitation Constraints}
%================================================================
\label{Sect:TighterBounds}

Besides the diffraction loss constraint, a less obvious (and competing) constraint exists, stemming from an abstract property of the eigenmodes of (\ref{eq:scaled_eigp}): {\em band limitation}.

%----------------------------------------------------------------
\subsubsection{Band-Limitation Property}
%----------------------------------------------------------------
Applying the HT operator at both sides of the eigenproblem in (\ref{eq:scaled_eigp}) and using the more or less obvious identities
\beq
{\cal H}_1[f] =
{\cal H}[\Pi({\bar r}) f({\bar r})],~~~
{\cal H}[{\cal H}[f]]=f,
\label{eq:ident}
\eeq
with $\Pi$ denoting the unit rectangular-window function, $\Pi(\xi)=1, 0\le\xi\le1$, $\Pi(\xi)=0, \xi>1$, one obtains
\beq
{\cal H}\left[\phi \exp\left(\imath V\right)\right](\pi N_D {\bar r})=\imath \frac{\pi N_D}{{\bar \gamma}}  \Pi\left({\bar r}\right)
\exp\left[-\imath V\left({\bar r}\right)\right]\phi\left({\bar r}\right).
\label{eq:bandlim}
\eeq 
Equation (\ref{eq:bandlim}) shows that the HT of the function $\phi \exp\left(\imath V\right)$ (and, {\em a-fortiori}, of the function $\phi$) has a compact support, vanishing outside $[0, \pi N_D]$. Technically, the HT plays the role of a wavenumber spectrum, and accordingly $\pi N_D$ is the {\em spatial bandwidth} of the field. Note that the spatial bandwidth is proportional to the number of electromagnetic degrees of freedom $N_D$ in (\ref{eq:EDF}).
It is therefore natural to try approximating  the optimal (but, as anticipated, unphysical) beam profiles obtained in Sect. \ref{Finite}  using a basis  in $L^2_{[0,\infty[}$ with finite spatial bandwidth $\pi N_D$. It is worth stressing that {\em no} constructive procedure is given for retrieving a mirror profile for which such a superposition is an actual eigenmode. Nonetheless, being a physically admissible (finite spatial bandwidth) profile, it is expected to yield {\em tighter noise} bounds, as compared to (\ref{eq:theoret}).

%----------------------------------------------------------------
\subsubsection{Prolate-Spheroidal Wave-Function Expansion}
\begin{figure}
\begin{center}
\includegraphics[width=8cm]{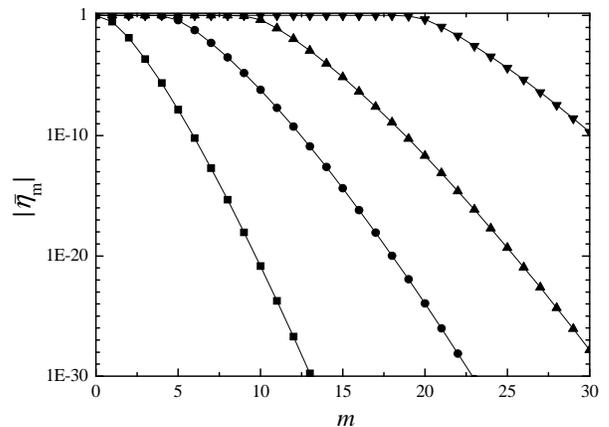}
\end{center}
\caption{PSWF eigenvalues (magnitude) as a function of order $m$, for various values of $N_D$. The semi-log scale highlights the step behavior with exponential tail (see the discussion in Sect. \ref{Sect:PSWF}). Squares: $N_D=1$; circles: $N_D=5$; up-triangles: $N_D=10$; down-triangles: $N_D=20$.}
\label{Figure3}
\end{figure}

%----------------------------------------------------------------
\label{Sect:PSWF}
A more or less obvious choice for the space-bandlimited basis is provided by the so-called {\em prolate-spheroidal wave-functions} (PSWFs) \cite{Slepian, HXiao, Shkolnisky2007} \footnote{Throughout the paper, this more generic denomination will be utilized, instead of the more detailed ``PSWFs on a disc'' in \cite{Shkolnisky2007}.}, which satisfy the eigenproblem \footnote{The PSWF basis is the spectrum of the eigenproblem in (\ref{eq:scaled_eigp}) with $V=0$, representing the eigenmodes of a confocal spherical Fabry-Perot cavity with finite-size mirrors.}
\beq
{\bar \eta} \varphi({\bar r})=\imath \pi N_D {\cal H}_1\left[\varphi\right](\pi N_D {\bar r}).
\label{eq:prolate1}
\eeq
In our implementation, the PSWFs are calculated following the approach in \cite{Shkolnisky2007}. 
It can be shown that (apart from irrelevant complex multiplicative constants) the solutions of (\ref{eq:prolate1}), $\varphi_n$, are real and satisfy the {\em double orthogonality} condition
\beq
\left<
\varphi_n,\varphi_m
\right>=\delta_{nm},~~
\left<
\varphi_n,\varphi_m
\right>_1={\bar \eta}_n \delta_{nm},
\label{eq:orthnorm}
\eeq
where $\delta_{mn}$ is the Kronecker symbol, ${\bar \eta}_n$ indicates the $n$-th eigenvalue of (\ref{eq:prolate1}), and
$\left<\cdot,\cdot \right>$ and 
$\left<\cdot,\cdot \right>_1$ denote the $L^2_{[0,\infty[}$ and $L^2_{[0,1]}$ (cylindrical) inner product, respectively.
The eigenvalue spectrum of (\ref{eq:prolate1}), shown in Fig. \ref{Figure3} for several values of $N_D$, has a {\em step-like} behavior: the first $\sim N_D$ eigenvalues are close to one in magnitude, while the remaining decay exponentially to zero \cite{HXiao}. The semi-log scale utilized in Fig. \ref{Figure3} highlights the step-like behavior (with exponentially-decaying tail) of the eigenvalue spectrum.  
%############################################################
%                Figure3
%
%###########################################################
The double-orthogonality condition in (\ref{eq:orthnorm}) implies 
\beq
\int_{1}^{\infty}
|\varphi_m({\bar r})|^2
{\bar r} d{\bar r} = 1 - |{\bar \eta}_m|^2.
\label{eq:diff_loss}
\eeq
%############################################################
%                Figure4
%
\begin{figure*}
\begin{center}
\includegraphics[width=12cm]{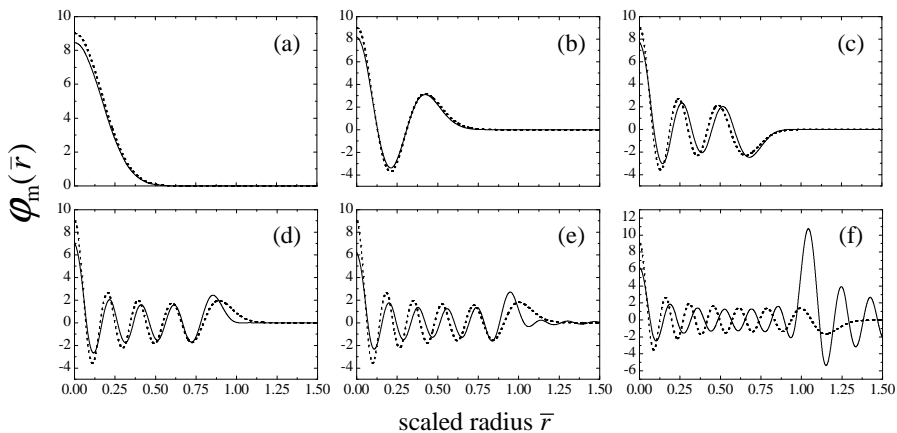}
\end{center}
\caption{PSWF profiles for $N_D=11.58$ and various orders (solid curves). (a): $m=0$; (b): $m=2$; (c): $m=5$; (d): $m=8$; (e): $m=10$; (f): $m=13$. Also shown, as reference (dashed curves), is the behavior of the infinite-mirror (GL-type) solutions.}
\label{Figure4}
\end{figure*}
%###########################################################
In view of (\ref{eq:diff_loss}) and the noted behavior of the eigenvalues, the first $\sim N_D$ eigenmodes are almost fully {\it localized} in $[0,1]$, while the remaining ones are almost fully (de-)localized to $\bar{r} > 1$. 
A plot of a few PSWFs of increasing order is shown in Fig. \ref{Figure4}, for $N_D=11.58$. Also shown, as a reference, is the behavior of the infinite-mirror (GL-type) solutions. It is observed that the agreement between the two is rather good for low orders, for which the functions are localized, and deteriorates as the order $m$ approaches $N_D$, beyond which the functions exhibit the anticipated de-localization.

The {\em best} (in $L^2$ norm) band-limited approximation of the compact-support minimum-noise beam profiles is therefore provided by the PSWF expansion
\beq
\phi_{BL}(\bar{r})=
\sum_{m=0}^{M_T-1}
c_m
\varphi_m(\bar{r}),
\label{eq:exp}
\eeq
with the coefficients $c_m$ obtained via Fourier-type projection \footnote{In our approach, based on the PSWF computation via the polynomial expansions in \cite{Shkolnisky2007}, the scalar products in (\ref{eq:cm}) can be efficiently computed in a semi-analytic fashion.},
\beq
c_m=\frac{\left<
\phi_s,\varphi_m
\right>_1}{\sqrt{\displaystyle{\sum_{n=0}^{M_T-1}} \left(\left<
\phi_s,\varphi_n
\right>_1\right)^2}}.
\label{eq:cm}
\eeq
%############################################################
%                Figure5
%
\begin{figure}
\begin{center}
\includegraphics[width=8cm]{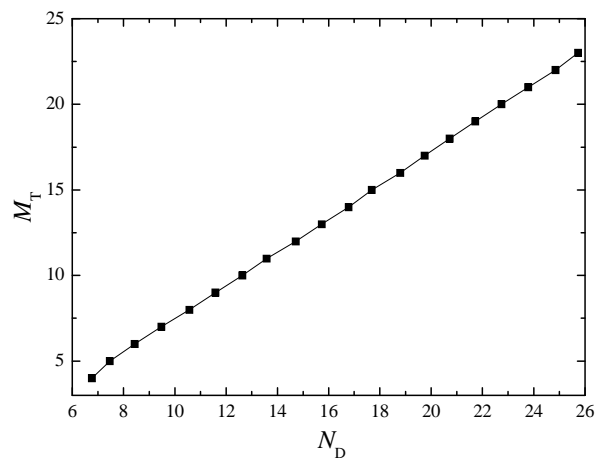}
\end{center}
\caption{Truncation index $M_T$ in the PSWF expansion (cf. (\ref{eq:MT})) as a function of number of electromagnetic degrees of freedom $N_D$.}
\label{Figure5}
\end{figure}
%###########################################################
It is readily shown that the truncation order $M_T$ in (\ref{eq:exp}) is dictated by the prescribed diffraction loss.
Under the ideal step-like assumption for the eigenvalue dependence on index,
whereby $|{\bar \eta}_m| = 1$, $\forall m \leq N_D$, and ${\bar \eta}_m=0$, $\forall m > N_D$, 
the diffraction-loss constraint would be satisfied for {\em any} $M_T \leq N_D$, 
{\em however small} the prescribed ${\cal L}_T$. 
A conservative estimate of $M_T$, taking into account the actual,
albeit tiny, departure of the $m< N_D$ eigenvalue magnitudes from unity may be obtained
from the obvious inequality
\begin{eqnarray}
{\cal L}\left[\phi_{BL}\right]&=&\sum_{m=0}^{M_T-1}
(1-|{\bar \eta}_m|^2)|c_m|^2\nonumber\\
&\leq&
(1-|{\bar \eta}_{M_T-1}|^2)\sum_{m=0}^{M_T-1}|c_m|^2\nonumber\\
&=&
(1-|{\bar\eta}_{M_T-1}|^2),
\label{eq:bnd}
\end{eqnarray}
where use has been made of the double-orthogonality conditions in (\ref{eq:orthnorm}), the fact that the $|{\bar \eta}_m|$
form a monotonically-decreasing sequence, and the unit-norm constraint, viz.
\beq
\sum_{m=0}^{M_T-1}|c_m|^2 =
\left\|\phi_{BL}\right\|^2 = 1.
\eeq 
We accordingly get the following (conservative) estimate
for the truncation order, which sets the {\it effective dimension} (number of
available design parameters) of the beam (mirror) optimization problem:
\beq
M_T = \mbox{largest } m : (1-|{\bar \eta}_{m-1}|^2) \leq {\cal L}_T.
\label{eq:MT}
\eeq
For ${\cal L}_T=1$ppm, the truncation index $M_T$ computed from (\ref{eq:bnd}) is plotted as a function of $N_D$ in Fig. \ref{Figure5}.  We may loosely conclude that the effective dimension of the optimization problem is 
\beq
M_T \lesssim N_D.
\label{eq:MTND}
\eeq 
%############################################################
%                Figure6
%
\begin{figure}
\begin{center}
\includegraphics[width=8cm]{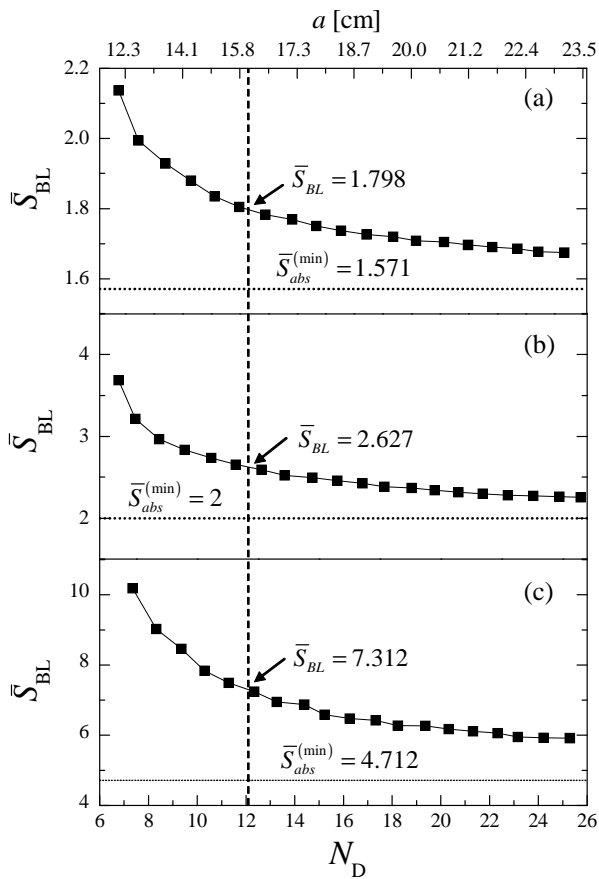}
\end{center}
\caption{Realistic noise bounds obtained from band-limited beam profiles (cf. (\ref{eq:exp})) and truncation index $M_T$ as in (\ref{eq:MT}) (cf. Fig. \ref{Figure5}), as a function of $N_D$. (a): Substrate Brownian ($q=-1$); (b): Coating ($q=0$); (c): Substrate thermoelastic ($q=1$);  Also shown, as references, are the corresponding mirror-radius scale (top axis, assuming $L=4$km and $\lambda=1064$nm), the absolute bounds (dotted lines, cf. (\ref{eq:theoret})), and the noise values for $a=16$cm (i.e., $N_D=12.03$).}
\label{Figure6}
\end{figure}
%###########################################################
The inequality in (\ref{eq:bnd}) will be reasonably tight when representing functions that are essentially localized within the unit-disc (mirror-confined beams), whose projection onto the de-localized eigenstates with $m \gtrsim N_D$ 
is negligibly small. 

More or less obviously, the accuracy of (\ref{eq:exp}) is strictly dependent on (and expected to increase with) the number of terms in
the truncated expansion. It makes therefore sense to check how close one can get to the optimal 
profiles for various meaningful values of $N_D$ (and hence, via (\ref{eq:MT}), $M_T$).
%############################################################
%                Figure7
%
\begin{figure}
\begin{center}
\includegraphics[width=8cm]{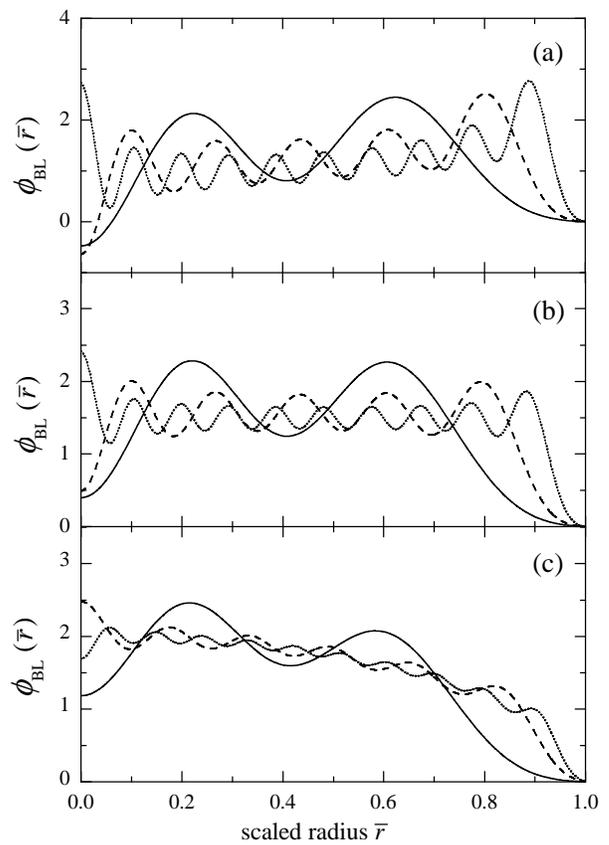}
\end{center}
\caption{As in Fig. \ref{Figure6}, but representative band-limited beam profiles for various values of $N_D$. Continuous curve: $N_D=6.77$ ($a=12$cm); dashed: $N_D=13.58$ ($a=17$cm); dotted: $N_D=22.74$ ($a=22$cm).}
\label{Figure7}
\end{figure}
%###########################################################

Figures \ref{Figure6}a-c show the behavior of the noises associated with the band-limited profiles in (\ref{eq:exp}), referred to as ${\bar S}_{BL}$, as a function of $N_D$. For all three cases, the noise decreases with increasing $N_D$, and appears to asymptotically approach values close to the absolute bounds in (\ref{eq:theoret}).
Recalling (\ref{eq:EDF}), the variation of $N_D$ was obtained by tuning the cavity length $L$ and the laser wavelength $\lambda$ at the reference values in Adv-LIGO, and taking the mirror radius $a$ within a realistic range (see the scale on the top axis of Fig. \ref{Figure6}). For each value of $a$ (and, hence, $N_D$), the truncation index $M_T$ (cf. Fig. \ref{Figure5}) was derived according to (\ref{eq:MT}), with ${\cal L}_T=1$ppm \footnote{In view of the inherent {\em discreteness} of the problem, it was not always possible to find values of $M_T$ so that the diffraction-loss constraint was tightly satisfied (i.e., diffraction loss smaller but close to the targeted value). In order to ensure the comparison among configurations yielding comparable diffraction losses, we slightly relaxed the constraint, looking for values of $M_T$ yielding diffraction losses {\em closest} ($\sim \pm 0.3$ppm) to the targeted value.}.  

Figure \ref{Figure7} shows the corresponding band-limited beam profiles for selected values of $N_D$. For the $q=0$ case (coating Brownian and thermoelastic noises), for instance, it is observed that, as $N_D$ increases, the profile tends to exhibit a more rapid ripple and a steepest decay. It can be argued that the {\em flatness} of the profile does not seem to be an essential ingredient for the coating noise reduction.   

To sum up, it is seen that the diffraction-loss constraint sets an upper limit to the
effective dimension of the optimization problem, via the finite-spatial bandwidth
property of the physically admissible solutions of the cavity eigenstate equation.
Thus, the {\em only}  way to approach the absolute minima of the noise constituents 
acting on the beam/mirror profiles, under a prescribed diffraction-loss constraint,
is by  increasing $N_D$, (i.e., if the cavity length and laser wavelength
are kept fixed, by increasing the mirror radius $a$).
However, as seen from Fig. \ref{Figure6}, increasing $N_D$ (aka, $a$) beyond a certain value
pays little, as the noise curves roll-off very slowly beyond a certain point, and tend to settle.
Going, e.g., from $a=16$ cm (Adv-LIGO baseline design) to $a=23$ cm reduces the
coating noise only by  $\sim 14\%$. Moreover, besides the technological challenges involved, this raises some model-consistency issues related to the actual validity of the underlying ITM approximation.

%----------------------------------------------------------------
\subsubsection{Validity of the ITM Approximation}
%----------------------------------------------------------------
\label{Sect:ITMValidity}
In \cite{Lovelace2006}, the ITM approximations in (\ref{eq:PSD}) have been validated and calibrated against the FTM numerical solutions in \cite{Agresti2005b}, for GB and MB profiles. Assuming a 40 kg fused-silica test mass and ${\cal L_T}=1$ppm (design specifications for Adv-LIGO), the ITM predictions for a MB profile were found to yield errors $<10\%$ in the coating (Brownian as well as thermoelastic) noises and $<25\%$ in the substrate Brownian noise (the thermoelastic noise component being negligible for fused-silica test-masses \cite{Agresti2005b}), for mirror radii $a\lesssim 17$cm. For sapphire test-masses, the error in the (dominant) substrate thermoelastic noise component was found to be comparable to the substrate Brownian case for fused silica. Assuming that comparable figures apply to the band-limited beam profiles in (\ref{eq:exp}) too, some representative values of the realistic bounds for the case $a=16$cm ($N_D=12.03$) are reported in Table \ref{Table2} (scaled to the corresponding absolute bounds in (\ref{eq:theoret})). One observes a moderate increase, as compared with the absolute bounds, of a factor $\sim 1.14$ for the substrate Brownian noise, $\sim 1.31$ for the coating (Brownian as well as thermoelastic) noises, and $\sim 1.55$ for the coating thermoelastic noise.

%%%%%%%%%%%%%%%%%%%%%%%%%%%%%%%%%%%%%%%%%%%%%%%%%%%%%%%%%%%%%%%%%
\section{Comparison with Current Status and Trends: Gaussian and ``Mesa'' Beams}
%%%%%%%%%%%%%%%%%%%%%%%%%%%%%%%%%%%%%%%%%%%%%%%%%%%%%%%%%%%%%%%%%
\label{Sect:current}
It is suggestive to compare the above derived bounds with those attainable by the current (Gaussian) and proposed (mesa) beam profiles.
For these profiles, without solving the eigenvalue problem in (\ref{eq:scaled_eigp}), 
one can exploit simple approximate analytic solutions for the dominant eigenmode, valid in the (transversely) infinite mirror limit, 
estimating the relevant diffraction losses via the so-called ``clipping approximation'' \cite{Dambrosio2003a}, 
i.e., by using the infinite-mirror approximate field distributions in
the first equation in (\ref{eq:diffloss1}).

%================================================================
\subsection{Gaussian Beams}
%================================================================
The scaled field distribution, in the infinite-radius-mirror approximation, for a GB can be expressed as
\beq
\phi_{GB}({\bar r},{\bar w}_0)=\Xi_{GB} \exp \left(
-\frac{{\bar r}^2}{2{\bar w}_0^2}
\right),
\eeq
where $\Xi_{GB}$ is a normalization constant, and the waist parameter ${\bar w}_0$ is fixed by the clipping-approximated diffraction-loss constraint,
\beq
{\bar w}_{0c}=\left(-\log{\cal L}_T\right)^{-\frac{1}{2}}.
\eeq
In view of the particularly simple analytic expression of the field intensity distribution (and of its HT), the scaled noise functional in (\ref{eq:caN_Dorm1}) can be computed in closed form,
\begin{eqnarray}
{\bar S}_{GB}
&\equiv&
\left\|
{\bar \kappa}^{\frac{q}{2}}~
{\cal H}\left[\left|\phi_{GB}\right|^2\right]
\right\|^2\nonumber\\
&=&
\frac{2^{\frac{q}{2}}  
\Gamma \left(\frac{q}{2}+1\right)
\left(-\log 
{\cal L}_T\right)^{\frac{q}{2}+1}
}{(1-{\cal L}_T)^2},~~q\ge-1.
\label{eq:GBN}
\end{eqnarray}

%================================================================
\subsection{Mesa Beams}
\begin{figure}
\begin{center}
\includegraphics[width=8cm]{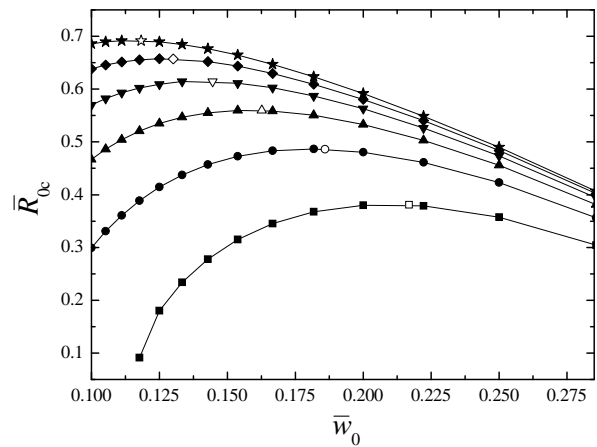}
\end{center}
\caption{Relationship between the MB parameters ${\bar R}_0$ and ${\bar w}_0$ arising from the clipping-approximated diffraction-loss constraint (${\cal L}_T=1$ppm), for various values of $N_D$ (assuming $L=4$km and $\lambda=1064$nm). Squares: $N_D=6.77$ ($a=12$cm); circles: $N_D=9.21$ ($a=14$cm); up-triangles: $N_D=12.03$ ($a=16$cm); down-triangles: $N_D=15.23$ ($a=18$cm); diamonds: $N_D=18.80$ ($a=20$cm); stars: $N_D=22.74$ ($a=22$cm). White bullets mark the minimum-spreading configurations (cf. (\ref{eq:MS})).}
\label{Figure8}
\end{figure}

\begin{figure}
\begin{center}
\includegraphics[width=8cm]{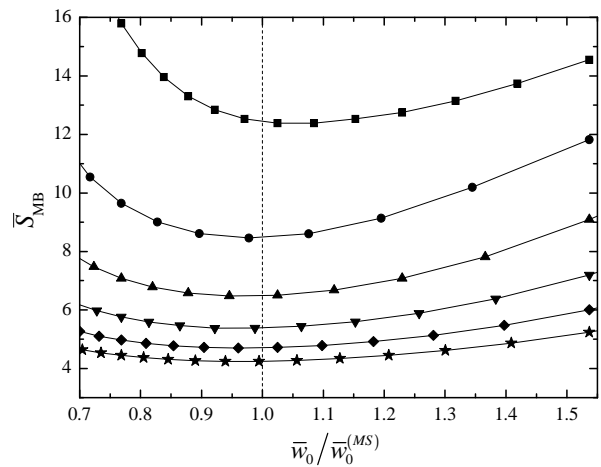}
\end{center}
\caption{As in Fig. \ref{Figure8}, but coating ($q=0$) noises as a function of ${\bar w}_0$ (scaled to its minimum-spreading value).}
\label{Figure9}
\end{figure}

%================================================================
A MB profile supported by a nearly-flat MH-shaped mirror can be synthesized via coherent superposition of GBs, with identical waist parameter $w_0$ and parallel optical axes, launched from a circular aperture of radius $R_0$ in the waist plane.
As shown in \cite{Galdi2006,Galdi2006a}, in the infinite-mirror approximation, such a beam profile can be effectively represented in terms of a Gauss-Laguerre (GL) expansion, which, in the scaled form used here, can be written as \footnote{A similar GL expansion was also derived for analytic parameterization of the MH-shaped mirror-deformation profile (cf. \cite[Eq. (34)]{Galdi2006a}).}
\begin{eqnarray}
\!\!\!\!\!\phi_{MB}({\bar r},{\bar R}_0,{\bar w}_0,N_D)\!\!&=&\!\!\Xi_{MB} \exp\left[\imath\Theta\left({\bar r}\right)\right]\nonumber\\
\!\!&\times&\!\!\sum_{m=0}^{\infty} A_m({\bar R}_0,{\bar w}_0,N_D)\nonumber\\
\!\!&\times&\!\!\!\psi_m\!\!\left(\frac{\sqrt{2}{\bar r}}{{\bar w}_0\sqrt{1+\frac{1}{\pi^2 N_D^2 {\bar w}_0^4}}}\right).
\label{eq:GLBE}
\end{eqnarray}
In (\ref{eq:GLBE}), $\Xi_{MB}$ is a normalization constant, $\Theta\left({\bar r}\right)$ is an irrelevant phase distribution, and the expansion coefficients $A_m$ are given by
\begin{eqnarray}
\!\!\!\!\!\!\!A_m({\bar R}_0,{\bar w}_0,N_D)\!\!\!&=&\!\!\!(-1)^m P\!\left(m+1,\frac{{\bar R}_0^2}{2{\bar w}_0^2}\right)\nonumber\\
\!\!\!&\times&\!\!\!\exp\!\left[
2\imath m \arctan\left(
\frac{1}{\pi N_D{\bar w}_0^2}
\right)
\right],
\label{eq:GL1}
\end{eqnarray}
with $P$ denoting an incomplete Gamma function \cite[Eq. (6.5.13)]{Abramowitz}. The 
orthonormal GL basis functions in (\ref{eq:GLBE}) are
\beq
\psi_m(\xi)=\sqrt{2}\exp\left(-\frac{\xi^2}{2}\right) L_m(\xi^2),
\label{eq:psim}
\eeq
with $L_m$ denoting an $m$th-order Laguerre polynomial \cite[Chap. 22]{Abramowitz}.
In the most general case, the scaled MB profile depends on three parameters: ${\bar R}_0$, ${\bar w}_0$, and $N_D$.
However, the (clipping-approximated) diffraction-loss constraint introduces a relationship ${\bar R}_0={\bar R}_{0c}\left({\bar w}_0,N_D\right)$, illustrated in Fig. \ref{Figure8} for several representative values of $N_D$, which reduces the number of independent parameters to two (${\bar w}_0, N_D$).
It is worth noticing that, in the topical MB literature, the waist parameter ${\bar w}_0$ is heuristically chosen according to a {\em minimum spreading} criterion \footnote{The Rayleigh distance of the GBs launched is chosen as coincident with the distance from the aperture plane to the mirror.}, viz.
\beq
{\bar w}_0^{(MS)}\equiv\frac{1}{a}\sqrt{\frac{k}{L}}=\frac{1}{\sqrt{N_D}},
\label{eq:MS}
\eeq
in an attempt of achieving the best tradeoff between {\em top-flatness} and {\em edge-steepness} of the beam intensity profile.   
This further reduces the number of independent parameters to one ($N_D$). 
When minimizing the noise functional in (\ref{eq:MinProb}), the above choice, while intuitively sound, is not necessarily justified {\em a priori} from the mathematical viewpoint, and the more general {\em two-parameter} optimization problem 
\beq
{\bar S}^{(min)}_{MB}= 
\min_{{\bar w}_0, N_D\in {\mathbb R}} 
\left\|
{\bar \kappa}^{\frac{q}{2}}~
{\cal H}\left[\left|\phi_{MB}\right|^2\right]
\right\|^2
\label{eq:SMB}
\eeq
should be considered instead. However, the heuristic minimum-spreading criterion in (\ref{eq:MS}) is pretty close to optimal for sufficiently large $N_D$.
As an illustrative example, the functional in (\ref{eq:SMB}) for $q=0$ (coating Brownian and thermoelastic noises) is shown in Fig. \ref{Figure9} as a function of ${\bar w}_0$ (scaled to the reference minimum-spreading value in (\ref{eq:MS})), for several representative values of $N_D$, within the parametric range of potential interest for Adv-LIGO. It is observed that the curves at fixed $N_D$ exhibit a broad minimum around (but not exactly at) ${\bar w}_0={\bar w}^{(MS)}_0$, which becomes {\em deeper} and {\em broader} as $N_D$ is increased. A similar behavior is observed for the substrate Brownian  ($q=-1$) and thermoelastic ($q=1$) noises,
and is not shown for brevity. 

%############################################################
%                Figure8
%
%###########################################################

%############################################################
%                Figure9
%
%###########################################################

%================================================================
\subsection{GB vs. MB vs. Absolute and Realistic Bounds}
%================================================================
The noise levels achievable via a GB profile and a reference MB profile (minimum noise, for $a=16$cm, i.e., $N_D=12.03$, cf. Fig. \ref{Figure9}), scaled to the corresponding absolute and realistic bounds derived in Sect. \ref{Sect:theoret}, are also included in Table \ref{Table2} \footnote{Note that, although the GB and MB results in Table \ref{Table2} have been obtained, for computational convenience, using the infinite-mirror approximation in (\ref{eq:GLBE}), they were checked to agree  to within $15\%$ with those obtained from the numerical solution (via the Nystrom method \cite{Canino}) of the eigenproblem in (\ref{eq:scaled_eigp}), for a number of representative cases.}. As already established in \cite{Agresti2005b}, MB profiles yield consistently lower noises than the GB counterparts, with reductions of nearly a factor $\sim 2.13$ in the
coating (Brownian as well as thermoelastic) noise, and of
$\sim 1.45$ and $\sim 3.07$ in the substrate Brownian and thermoelastic noise components, respectively. 

By comparison with the absolute and realistic bounds, one notes a potential for significant further reductions. Specifically, as compared to the MB reference values, the realistic bounds indicate potential reductions of nearly a factor $1.8$, $2.5$, and $2.9$ for the
substrate Brownian, coating (Brownian as well as thermoelastic) and substrate thermoelastic noises, respectively, thereby justifying the further exploration of alternative numerical-optimization-driven configurations. 

%%%%%%%%%%%%%%%%%%%%%%%%%%%%%%%%%%%%%%%%%%%%%%%%%%%%%%%
\begin{table*}
\caption{Comparison between absolute (cf. (\ref{eq:theoret})) and realistic (for $a=16$cm, i.e., $N_D=12.03$, estracted from Fig. \ref{Figure6}) noise bounds. Also shown, as references, are the noise levels attainable with GB and reference MB (minimum noise, for $a=16$cm, i.e., $N_D=12.03$, cf. Fig. \ref{Figure9}) profiles.}
\begin{ruledtabular}
\begin{tabular}{ccccccc}
$q$  & ${\bar S}_{abs}^{(min)}$ & ${\bar S}_{BL}/{\bar S}_{abs}^{(min)}$ & ${\bar S}_{GB}/{\bar S}_{abs}^{(min)}$ & ${\bar S}^{(min)}_{MB}/{\bar S}_{abs}^{(min)}$ & ${\bar S}_{GB}/{\bar S}_{BL}$ & ${\bar S}^{(min)}_{MB}/{\bar S}_{BL}$  \\
\hline
-1 & 1.5708 & 1.145 & 2.965 & 2.043 & 2.591 & 1.785\\
0 & 2 & 1.313 & 6.907 & 3.238 & 5.256 & 2.465\\
1 & 4.712 & 1.552 & 13.658 & 4.454 & 8.801 & 2.870\\
\end{tabular}
\end{ruledtabular}
\label{Table2}
\end{table*}
%%%%%%%%%%%%%%%%%%%%%%%%%%%%%%%%%%%%%%%%%%%%%%%%%%%%%%%

%================================================================
\subsection{Optimal vs. Good  Profiles}
%================================================================

On the basis of the above analysis, a robust (e.g., genetic) optimization  algorithm based on a mirror parameterization consistent with the problem's effective dimension, aimed at getting as close as possible to the realistic (lower) noise bounds, could be implemented with relative ease.

It should stressed, however, that any {\em optimal} design should also cope with some more or less obvious additional requirements to be also rated as a {\em good} design, e.g.,
\begin{itemize}

\item{The optimal mirror should note pose critical technological challenges;}

\item{The optimal field should be easy to launch, i.e., should couple effectively to the injected laser beam;}

\item{The optimal field should be reasonably stable against misalignment and mirror manufacturing tolerances.}

\end{itemize}

Any candidate sub-optimal designs should be ultimately gauged on the basis of their compliance with the above practical requirements.

%%%%%%%%%%%%%%%%%%%%%%%%%%%%%%%%%%%%%%%%%%%%%%%%%%%%%%%%%%%%%%%%%
\section{Conclusions and Perspectives}
%%%%%%%%%%%%%%%%%%%%%%%%%%%%%%%%%%%%%%%%%%%%%%%%%%%%%%%%%%%%%%%%%
\label{Conclusions}
In this paper, based on the ITM approximations in \cite{OShaug2006,Lovelace2006}, we have addressed some key preliminary issues in connection with the optimal beam-shaping problem for thermal noise reduction in advanced gravitational wave interferometric detectors. The main conclusions can be summarized as follows:

\begin{itemize}
	\item[{\em i)}] The estimated lower-bounds in a realistic configuration, accounting for physical-feasibility-induced (diffraction losses, band-limitation) and model-consistency (ITM approximation) constraints, indicate the possibility of significant noise reductions (cf. Table \ref{Table2}) as compared with the current status and trend. In particular, for the coating noise (dominant for the case of fused-silica test-masses), a potential reduction of nearly a factor 2.5 is estimated, as compared with the MB counterpart.

\item[{\em ii)}] The key role of the number $N_D$ of electromagnetic degrees of freedom (aka, the Fresnel number) of the optical cavity in establishing realistic lower-bounds has been highlighted. In this connection, while the possibility of increasing $N_D$ by acting on the cavity length or the laser wavelength does not appear technologically viable for second-generation detectors (thereby leaving the mirror radius as the only adjustable design parameter), it could be taken into account for third-generation instruments.

\item[{\em iii)}] From inspection of the band-limited beam profiles derived in Sect. \ref{Sect:PSWF}, one can infer important (sometime counter-intuitive) {\em prior information} to intelligently drive the optimization process. For instance, for the coating noises, it clearly emerges that the flatness of the beam is not a critical requirement, since profiles with ripples (cf. Fig. \ref{Figure7}) can perform better than flat-top MBs. This observation, which is also consistent with the results obtained using higher-order modes in spherical-mirror cavities \cite{Vinet2}, should be taken into account when parameterizing the functional space chosen for the optimization problem.   

\end{itemize}

We believe that the above results pave the way for the actual optimization problem, for which a genetic-algorithm \cite{Rahmat} implementation is currently under investigation. Interesting research directions include
 extensions of the preliminary study to higher-order (multipolar, non-axisymmetric) modes.

While proofreading this paper, we became aware of work done independently by 
M. Bondarescu  and Y. B. Chen \cite[Chap. 3]{BondarescuThesis}.
Using the GL basis \cite{Siegman} to synthesize the cavity field, 
they succeeded in retrieving a special mirror profile which minimizes 
the coating  (Brownian and thermoelastic) noise. Remarkably, their results
provide a nice independent confirmation of the general conclusions drawn here.
Indeed, they confirm the key role of the mirror radius 
(aka, cavity Fresnel number)  in setting the tradeoff  between 
diffraction loss and  noise reduction, in complete agreement 
with the general scenario outlined here. 
Also, the minimum noise achieved by their design gets pretty close 
to our corresponding realistic bound.

\begin{acknowledgments}
This material is based upon work supported by the US National Science Foundation under Cooperative Agreement No. PHY-0107417.
\end{acknowledgments}

\appendix

%%%%%%%%%%%%%%%%%%%%%%%%%%%%%%%%%%%%%%%%%%%%%%%%%%%%%%%%%%%%%%%%%
\section{Minimization of the Lagrangian Functional in (\ref{eq:lagrangian})}
%%%%%%%%%%%%%%%%%%%%%%%%%%%%%%%%%%%%%%%%%%%%%%%%%%%%%%%%%%%%%%%%%
\label{AppA}
The stationary solution of the variational problem obtained equating to zero the functional derivative of (\ref{eq:lagrangian}) can be easily derived using the Gateaux differential
\beq
\lim_{\epsilon\rightarrow0}\frac{\Lambda[f_s+\epsilon\delta\!f,\mu]-\Lambda[f_s,\mu]}{\epsilon}=0,~~\forall\delta\!f\in L^1_{[0,1]}.
\label{eq:Frechet}
\eeq
It is readily verified that   
\begin{eqnarray}
\!\!\!\!\Lambda[f+\epsilon\delta\!f,\mu]&=&\Lambda[f,\mu]
\!+\!
2\epsilon\left[
\left<{\bar \kappa}^{\frac{q}{2}}{\cal H}_1\left[f\right],
{\bar \kappa}^{\frac{q}{2}}{\cal H}_1\left[\delta\!f\right]
\right>\right.\nonumber\\
\!\!\!&-&\!\!\!\left.\mu \!\int_0^1 d{\bar r} {\bar r} \delta\!f\left({\bar r}\right)\right]
\!\!+\epsilon^2
\left\|
{\bar \kappa}^{\frac{q}{2}}~
{\cal H}_1\left[\delta\!f\right]
\right\|^2\!.
\label{eq:exLag}
\end{eqnarray}
By using (\ref{eq:exLag}) in (\ref{eq:Frechet}), and interchanging the spectral and spatial (HT) integrals in the inner product, the stationarity condition can be equivalently written as
\begin{eqnarray}
\int_{0}^1 \!\!\!\!\!\!\!\!&&d{\bar r} 
\left[
\int_{0}^\infty  d{\bar\kappa} {\bar\kappa}^{q+1} J_0({\bar\kappa}{\bar r})\right.\nonumber\\
&\times&\left.\!\!\!
\int_{0}^1  d{\bar r}'  {\bar r}' f_s({\bar r}')  J_0({\bar\kappa} {\bar r}')-\mu\right] 
{\bar r}\delta\!f({\bar r}) 
=0,\nonumber\\
&~&\forall\delta\!f\in L^1_{[0,1]},
\label{eq:statio1}
\end{eqnarray} 
from which it follows that the stationary profile $f_s$ satisfies the integral equation
\beq
\int_{0}^\infty  d{\bar\kappa} {\bar\kappa}^{q+1} J_0({\bar\kappa}{\bar r})
\int_{0}^1  d{\bar r}'  {\bar r}' f_s({\bar r}')  J_0({\bar\kappa} {\bar r}')=\mu,~~0\le{\bar r}\le 1.
\label{eq:inteq}
\eeq
Equation (\ref{eq:inteq}) can be solved in closed-form, by inspection. Indeed, we capitalize on its nested-HT structure to use the following integral identities  
\begin{eqnarray}
\int_{0}^\infty {\bar\kappa}^{\frac{q}{2}}
J_{\frac{q}{2}+1}({\bar\kappa}) J_0({\bar r}{\bar\kappa})
d{\bar\kappa}=2^{\frac{q}{2}} \Gamma\left(\frac{q}{2}+1\right),\nonumber\\
0\le{\bar r}\le 1,~-1\le q\le 1,
\label{eq:Prud1}
\end{eqnarray}
(see, e.g., \cite[Eq. (2.12.31.1)]{Prudnikov}), and
\begin{eqnarray}
\!\!\!{\bar\kappa}^{\frac{q}{2}}J_{\frac{q}{2}+1}({\bar\kappa})\!=\!
\frac{2^{-\frac{q}{2}} {\bar\kappa}^{q+1}}{\Gamma\left(\frac{q}{2}+1\right)} 
\int_{0}^1 d{\bar r}' {\bar r}' (1-{\bar r}'^2)^{\frac{q}{2}}J_0({\bar\kappa}{\bar r}'),\nonumber\\
q\ge-1,
\label{eq:Prud2}
\end{eqnarray}
(ibid, \cite[Eq. (2.12.3.6)]{Prudnikov}).
By combining (\ref{eq:Prud2}) and (\ref{eq:Prud1}) one obtains
\begin{eqnarray}
\frac{2^{-q}}{\left[\Gamma\left(\frac{q}{2}+1\right)\right]^2} 
\int_{0}^\infty d{\bar\kappa} {\bar\kappa}^{q+1} 
J_0({\bar\kappa}{\bar r})\nonumber\\
\times\int_{0}^1 d{\bar r}' 
{\bar r}'(1-{\bar r}'^2)^{\frac{q}{2}}
J_0({\bar\kappa}{\bar r}')=1,\nonumber\\
0\le{\bar r}\le 1,
\end{eqnarray}
which, by comparison with (\ref{eq:inteq}), yields
\beq
f_s({\bar r})=\mu\frac{(1-{\bar r}^2)^{\frac{q}{2}}}
{2^{q}\left[\Gamma\left(\frac{q}{2}+1\right)\right]^2},~~0\le{\bar r}\le 1.
\label{eq:sol}
\eeq
The as yet unspecified multiplier $\mu$ in (\ref{eq:sol}) is chosen so as to satisfy the normalization constraint in (\ref{eq:const2}),
\beq
\mu=2^q(q+2)\left[\Gamma\left(\frac{q}{2}+1\right)\right]^2,
\eeq
thereby yielding the final result in (\ref{eq:fs}).

From (\ref{eq:exLag}) and (\ref{eq:inteq}), it follows that  
\begin{eqnarray}
{\bar S}[g]-{\bar S}[f_s]=\left\|
{\bar \kappa}^{\frac{q}{2}}~
{\cal H}_1\left[g-f_s\right]
\right\|^2,\nonumber\\
\forall g\in L^1_{[0,1]}:\int_0^1 g\left({\bar r}\right) {\bar r} d{\bar r}=1.
\label{eq:SS}
\end{eqnarray}
Therefore, from the readily verifiable positive-definiteness of the functional norm in (\ref{eq:SS}),
one concludes that that the stationary profile $f_s$ in (\ref{eq:fs}) yields the {\em absolute minimum} of the noise functional in (\ref{eq:probcan}).

\end{document}